\documentclass[superscriptaddress,twocolumn, amsmath,amssymb, aps,prx,citeautoscript]{revtex4-2}
\setcitestyle{super}

\usepackage{graphicx}
\usepackage{booktabs}
\usepackage{siunitx}
\usepackage{amsfonts}
\usepackage{amsmath}
\usepackage{bm}
\usepackage{caption}
\usepackage{subcaption}
\usepackage[capitalize]{cleveref}

\begin{document}

\title{Hyperspectral Compressive Wavefront Sensing}

\author{Sunny Howard}
\affiliation{Department of Physics, Clarendon Laboratory, University of Oxford, Parks Road, Oxford OX1 3PU, United Kingdom}%
\affiliation{Ludwig--Maximilians--Universit{\"a}t M{\"u}nchen, Am Coulombwall 1, 85748 Garching, Germany}%

\author{Jannik Esslinger}
\affiliation{Ludwig--Maximilians--Universit{\"a}t M{\"u}nchen, Am Coulombwall 1, 85748 Garching, Germany}%

\author{Robin H.W. Wang}
\affiliation{Department of Physics, Clarendon Laboratory, University of Oxford, Parks Road, Oxford OX1 3PU, United Kingdom}%
\author{Peter Norreys}
\affiliation{Department of Physics, Clarendon Laboratory, University of Oxford, Parks Road, Oxford OX1 3PU, United Kingdom}%
\affiliation{John Adams Institute for Accelerator Science, Denys Wilkinson Building, Oxford OX1 3RH, United Kingdom}%
\author{Andreas Döpp}
\affiliation{Department of Physics, Clarendon Laboratory, University of Oxford, Parks Road, Oxford OX1 3PU, United Kingdom}%
\affiliation{Ludwig--Maximilians--Universit{\"a}t M{\"u}nchen, Am Coulombwall 1, 85748 Garching, Germany}%
\email{a.doepp@lmu.de}

\begin{abstract}
Presented is a novel way to combine snapshot compressive imaging and lateral shearing interferometry in order to capture the spatio-spectral phase of an ultrashort laser pulse in a single shot.  A deep unrolling algorithm is utilised for the snapshot compressive imaging reconstruction due to its parameter efficiency and superior speed relative to other methods, potentially allowing for online reconstruction. The algorithm's regularisation term is represented using neural network with 3D convolutional layers, to exploit the spatio-spectral correlations that exist in laser wavefronts. Compressed sensing is not typically applied to modulated signals, but we demonstrate its success here. Furthermore, we train a neural network to predict the wavefronts from a lateral shearing interferogram in terms of Zernike polynomials, which again increases the speed of our technique without sacrificing fidelity. This method is supported with simulation-based results. While applied to the example of lateral shearing interferometry, the methods presented here are generally applicable to a wide range of signals, including Shack-Hartmann-type sensors. The results may be of interest beyond the context of laser wavefront characterization, including within quantitative phase imaging.
\end{abstract}

\maketitle

\section{Introduction}
Ultrashort laser pulses possess necessarily broad spectral bandwidth\cite{jolly}. The chromatic properties of the optical elements that are used for generation or application of such pulses can then create relations between the spatial and temporal profiles, called spatio-temporal couplings (STCs) \cite{stcs_laser}. These phenomena can lead to a variety of effects including, for example, the broadening of a focused laser pulse either spatially or temporally, thereby reducing it's peak intensity\cite{peakduration}. Deliberately introduced STCs can lead to exotic light pulses that behave very differently from 'normal' pulses. Examples of this are the so-called flying focus\cite{flyingfocus} with its potential application in laser-driven wakefield accelerators\cite{wakefield} or orbital angular momentum beams\cite{ramythesis}. Universally, the expansion in the  applications of ultrafast laser pulses has exacerbated the need for a robust way to measure their properties. 

To resolve STCs, one must gain wavefront information over the three-dimensional hypercube ($x,y,t$) or equivalently its spatio-spectral analogue ($x,y,\omega$). Due to the limitation that array sensors (such as CMOS cameras) capture information in a maximum of two dimensions, the majority of current techniques resort to scanning over one or two dimensions; whether its a spatial \cite{seatadpole,miguel}, spectral\cite{hamster,FALCON} or temporal\cite{termites} scan. Such techniques are time consuming and are blind to shot-to-shot variations and drift of the laser. While there exist some methods that are single-shot\cite{stripefish} - i.e. those that capture the hypercube in one shot - these currently lack resolution, spectral range and are cumbersome to implement. 

Inspired by recent progress in machine-learning based laser science\cite{dopp2022data}, here we present the concept for a single-shot method, which utilises compressed sensing to resolve the wavefront in both the spectral and spatial domains. The paper is structured as follows. In Section II we will discuss the wavefront sensor and in Section III we introduce snapshot compressive imaging as a way to expand the wavefront sensor to measuring multiple colors at once. Our implementation is based on deep unrolling, which yields high performance in both reconstruction fidelity and speed, as required for use as a real-time diagnostic. Section IV lists a thorough description of all neural network architectures used, and Section V contains a description of how training data was generated, before Section VI displays the results of the proposed method.

\section{Wavefront sensing}\label{wavefront}
\begin{figure*}
    \centering
    \includegraphics[width=17.5cm]{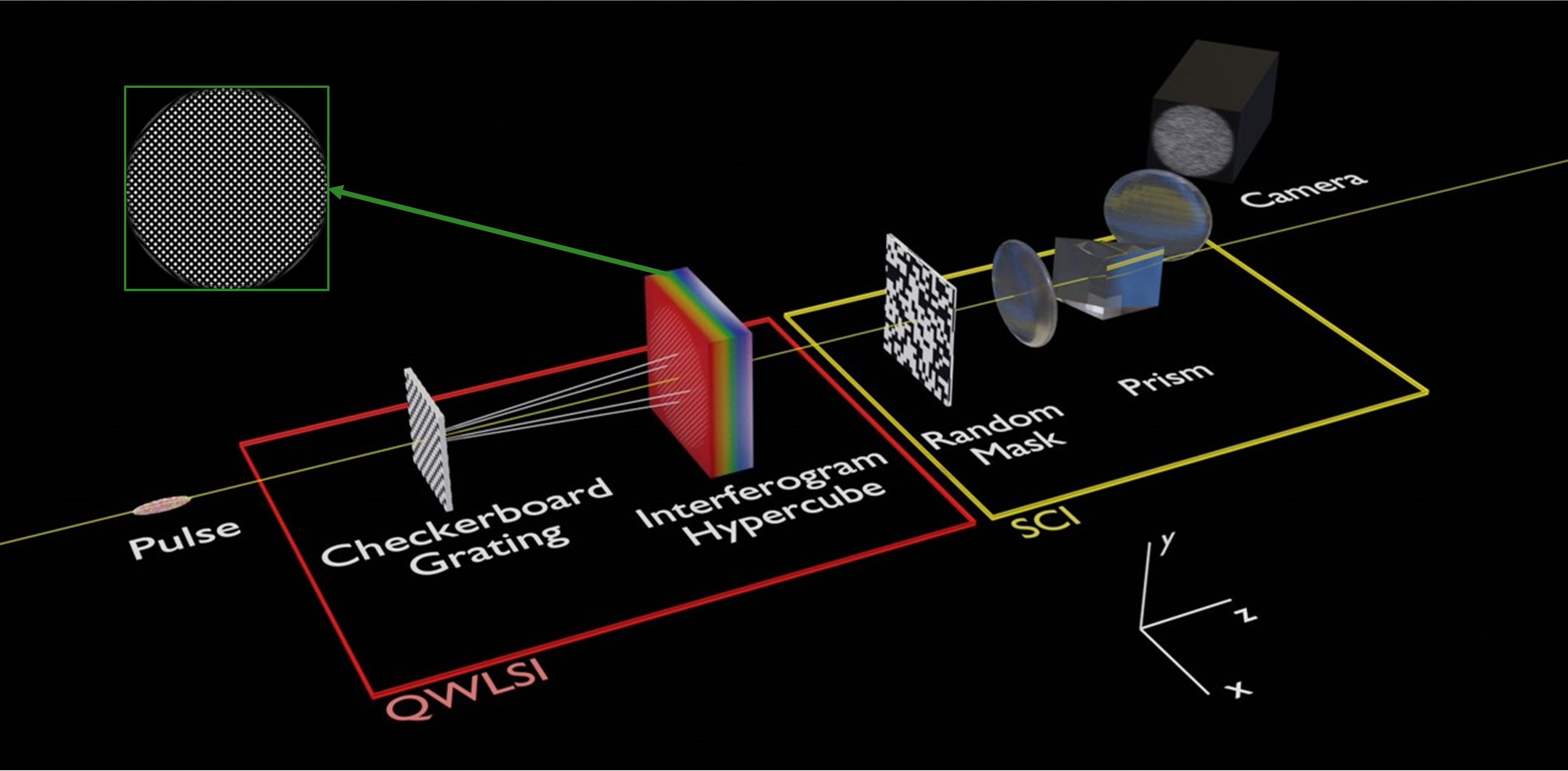}
    \caption{Schematic of the experimental setup that was simulated. The pulse first travels through a quadriwave lateral shearing interferometer, yielding a hypercube of interferograms, a slice of which is shown in the green box. The hypercube is then passed through a CASSI setup. This consists of a random mask and a relay system encompassing a prism, before the coded shot is captured on the camera. This diagram is not to scale.}
    \label{fig:setup}
\end{figure*}
The wavefront sensor that was simulated in this example was a Quadriwave Lateral Shearing Interferometer, which is known for its high resolution and reconstruction fidelity. Nonetheless, our method can in general be applied to any kind of wavefront retrieval technique, including the popular Shack-Hartmann sensor or multi-plane techniques such as Gerchberg-Saxton phase retrieval. 

A Lateral shearing interferometer (LSI) measures the spatially varying phase of a light beam, and was first applied to the measurement of ultrashort laser pulses in the late 1990s \cite{lsi_laser}. The LSI works by creating multiple copies of the laser pulse and shearing them laterally relative to each other before their interference pattern is captured on a sensor. Due to the shear, information about the spatial gradient of the wavefront is encoded in the interferogram. This can then be extracted using Fourier filtering \cite{fourierfilter} and stitched together to form the wavefront via methods such as modal reconstruction\cite{modal} or Fourier integration \cite{fourierintegration}. The most popular implementation is the aforementioned  quadriwave lateral shearing interferometer (QWLSI). By generating and shearing four ($2\times 2$) copies of the pulse under investigation, this setup also enables the extraction of two separate pairs of orthogonal gradients, meaning two distinct estimates for the wavefront can be found, providing error estimation. This property is highly desirable for a sensor based on compressed sensing, because inevitable noise in the measurement can corrupt the wavefront with reconstruction artifacts. This would for instance be the case in a two-plane Gerchberg-Saxton algorithm. In contrast, redundancy of phase information in QWLSI provides direct validation and thus, makes the wavefront retrieval much more resilient to noise. A sketch illustration the concept of QWLSI is shown in the red box of \cref{fig:setup}.

\subsection{QWLSI Simulation}\label{QWLSI_Simulation}
A physical implementation of QWLSI usually consists of a phase grating with `pixels' of alternating phase arranged in a checkerboard pattern \cite{quadri2}, which leads to dominant diffraction in $2\times 2$ copies of the beam. Instead of simulating this process, we consider an idealized setup where we analytically generate the four copies. We begin with creating the pulse of interest by defining a spatial-spectral intensity, $I_0$ and phase, $\phi_{0}$,
\begin{equation}
    E_{0}(x,y,z=0,\omega) = \sqrt{I_0(x,y,z=0,\omega)} e^{i\phi_{0} (x,y,z=0,\omega)}.
\end{equation}
The pulse is copied 4 times, and each copy's field is propagated to the detector plane according to the following rules. Note that for brevity, when the $z$ index is not stated, $z=0$ and the $\omega$ index will be suppressed from the electric field.
\subsubsection{Propagation}
Considering the $j^{th}$ copy has travelled a distance $\Delta z$, its poloidal angle is $\theta_{j}$ and its azimuthal angle is $\zeta$, then one finds its displacement in the $x$ and $y$ directions to be:
\begin{align*}
    \Delta x_{j}(\omega) &= \Delta z\sin(\theta_{j})\sin(\zeta(\omega)),\\
    \Delta y_{j}(\omega) &= \Delta z\cos(\theta_{j})\sin(\zeta(\omega)).
\end{align*}

In a QWLSI, the poloidal angles are $\theta_{j}\in[0,\frac{\pi}{2},\pi,\frac{3\pi}{2}]$. In \cref{fig:setup}, one identifies the azimuthal angle, $\zeta$ as that between the white copy lines, and the central yellow line. This is related to both the pitch of the grating $\Lambda$ and the wavelength $\lambda$ by:
\begin{equation*}
    \zeta(\omega)=\arcsin{\left(2\pi\frac{\lambda}{\Lambda}\right)}.
\end{equation*}

The resulting electric field of the copy is:

\begin{equation}
    E_j(x,y,\Delta z) = \frac{1}{4}\sqrt{I_{0}(x-\Delta x_{j},y-\Delta y_{j})} e^{i\phi_{0}(x-\Delta x_{j},y-\Delta y_{j})}.
\end{equation}

\subsubsection{Tilt}
As diffraction occurs at an angle $\zeta$, the grating imparts a tilt onto the copy. This translates to an additional phase shift dependant on both the spectral and spatial domains,

\begin{equation*}
    \Delta\phi_{j}(x,y,\omega)  = k(\omega)\left(x \cos(\theta_{j}) + y \sin(\theta_{j})\right ) \sin(\zeta),
\end{equation*}

where $k=\frac{2\pi}{\lambda}$ is the wavevector of the pulse. This tilt is crucial in reconstruction as it provides a high frequency modulation that separates the gradients in Fourier space.

Combining these two effects and summing over copies, we obtain the final changes to the field of: 
\begin{align}
\begin{split}
        E(x,y,\Delta z) = \frac{1}{4}\sum_{j=1} ^{4}  &\sqrt{I_{0}(x-\Delta x_{j},y-\Delta y_{j})}  \\
        &\cdot e^{i\left(\phi(x-\Delta x_{j},y-\Delta y_{j}) + \Delta\phi_{j}(x,y) \right)}
\end{split}
\end{align}
 At the Talbot self-imaging plane, $ \Delta z= {2\Lambda^{2}}/{\lambda }$, one has a hypercube of interferograms. An example of a one frequency channel slice is shown in the green box of \cref{fig:setup}.

 In other applications one would collapse the cube onto a sensor at this point; however this would eliminate the chance of retrieving the spectrally-resolved phase. Instead, as discussed in Section III, we use snapshot compressive imaging to aid in the capturing of the cube.

\subsection{Wavefront Reconstruction}

Once the interferogram is captured, one must extract the wavefront. As previously mentioned, current reconstruction methods usually involve multiple steps, i.e. extracting the gradients, integrating and stitching them together. This can be a time consuming process, especially in a hyperspectral setting where the reconstruction has to be done for every channel. To address this problem, we present a deep learning approach to wavefront reconstruction for LSI. While similar work has been done in the context of Shack Hartmann sensors \cite{shackhart1,shackhart2}, this is the first application of deep learning to LSI reconstruction, to the best of our knowledge. The network that was used will be discussed in Section IV.

\section{Snapshot Compressive Imaging}\label{sci}

Compressed sensing (CS) describes the highly efficient acquisition of a sparse signal from less samples than would classically be required according to the Nyquist theorem, by utilising optimisation methods to solve underdetermined equations. Snapshot compressive imaging (SCI) is an example of CS, capturing three dimensional data on a two dimensional sensor in a single shot. 

Fundamentally, there are two requirements to be fulfilled for CS to work. First, the signal must be sparse in some basis, and second, the signal must be sampled in a basis that is incoherent with respect to the sparse basis  \cite{candes}. The first condition was hypothesised to be satisfied given the fact that laser wavefronts are known to be well-expressed with a few coefficients of the Zernike basis. When one doesn't have prior knowledge about which basis the signal is sparse in, the second condition is often solved by performing random sampling. Whilst being trivial for two dimensional data, in the context of SCI it is challenging, as the 3D hypercube must be randomly sampled onto a 2D sensor. To do so, nearly all research in this area uses hardware based on the coded aperture snapshot compressive imaging (CASSI) system \cite{cassi1,cassi2}. 
\subsection{CASSI}
The hypercube is first imaged onto a coded aperture. This is a binary random mask with each pixel transmitting either 100\% or 0\% of the light. The cube is then also passed through some dispersive medium, e.g. a prism or grating, before being captured by a sensor resulting in what is known as the \textit{coded shot}. The effect of this optical system is that when the hypercube reaches the detector plane, each spectral channel is encoded with a different coded aperture, thereby approximating random sampling across the whole cube. It is then possible for a reconstruction algorithm to retrieve the cube. A diagram of a CASSI system is shown in the yellow box in \cref{fig:setup}, with an example of a coded shot for an interferogram hypercube shown on the far left of \cref{fig:finalresult}. The setup can easily be simulated by multiplying the cube by the mask, then shifting the channels of the cube according to the amount of (angular) dispersion imparted onto them, and finally summing over the spectral axis. 

Mathematically, the CASSI system discussed above is summarized into a matrix $\bm{\Phi}$, which operates on $\bm{m}$, a vectorized representation of the hypercube, to give $\bm{n}$, a vectorized coded shot,

\begin{equation}
    \bm{n} = \bm{\Phi}\bm{m}.
\end{equation}

In order to reconstruct $\bm{m}$, one can solve
\begin{equation}
    \tilde{\bm{m}} = \textrm{argmin}_{\bm{m}}\left[ \underbrace{||\bm{n} - \bm{\Phi}\bm{m}||^{2}}_{\mbox{\small{data term}}} + \underbrace{\eta \mathcal{R}(\bm{m},\psi)}_{\mbox{\small{regularizer}}}\right].\label{argmin}
\end{equation}
 The first term on the right hand side is labelled the data term, and enforces that the hypercube must match the coded shot when captured. This alone would be an under-determined system, so a regularization term, parameterised by $\psi$, is added which restricts the solution space and selects the correct hypercube. 

Most methods that have been developed to solve this non-convex equation can be sorted into two classes: iterative algorithms or end-to-end neural networks. The former offers good generalisation but lacks abstraction capability and is slow, whilst deep nets are fast and have been shown to learn almost any function, but can be prone to overfitting \cite{snapshotcompressiveimaging}. A middle ground that offers state of the art performance is deep unrolling.

\subsection{Deep Unrolling}
While an end-to-end neural net would attempt to solve \cref{argmin} directly, 
if it were possible to split the equation, the data term can actually be solved analytically. This is desirable as it alleviates the abstraction needed to be done by the network resulting in greater generalisation and better parameter efficiency \cite{algorithm_unrolling}. To perform such a separation, half quadratic splitting is employed. First an auxiliary variable $\bm{p}$ is substituted into the regularization term, with equation 6 being equivalent to equation \ref{argmin}. Then, the constraint is relaxed and replaced by a quadratic loss term,
\begin{align}
    \bm{\hat{m}},\bm{\hat{p}} &= \textrm{argmin}_{m,p} \left[|\bm{n} - \bm{\Phi} \bm{m}|^{2}  +\eta R(\bm{p})\right]  \:\: \textrm{s.t.} \: \bm{m}=\bm{p},\\
    &\approx \textrm{argmin}_{m,p} \left[|\bm{n} - \bm{\Phi} \bm{m}|^{2}  +\eta R(\bm{p}) + \beta |\bm{m}-\bm{p}|^{2} \right].\label{minimizeboth}
\end{align}
Here, $\beta$ is a variable that controls the strength of the constraint. High values of $\beta$ will strongly enforce $\bm{m} = \bm{p}$ and approximate the subject-to statement. 

The benefit of this problem formulation is that it is then possible to split \cref{minimizeboth} into two minimization sub-problems in $\bm{m}$ and $\bm{p}$, and effectively separate the data term from the regularization term. When minimized iteratively, the following sub-problems can approximate  \cref{minimizeboth}:

\begin{equation}
    \bm{\hat{p}}^{k+1} = \textrm{argmin}_p \left[\beta|\bm{p} - \bm{m}^{k}|^{2} + \eta R(\bm{p})\right] \sim \mathcal{S}(\bm{m}^{k}),
    \label{secondsub}
\end{equation}
\begin{equation}
    \bm{\hat{m}}^{k+1} = \textrm{argmin}_m \left[|\bm{n} - \bm{\Phi} \bm{m}|^{2} + \beta|\bm{p}^{k+1} - \bm{m}|^{2} \right]. \label{firstsub}
\end{equation}

\cref{firstsub} is a convex equation and can be solved via a conjugate gradient algorithm which provides better stability than solving analytically. $\mathcal{S}$ on the right-hand side of \cref{secondsub} represents that a neural network will be used to solve the equation. 

The deep unrolling process is shown in \cref{fig:networks}(bi). Firstly $\bm{m}^{(0)}$ is initialized: $\bm{m}^{(0)} = \bm{\Phi}^{T}\bm{n}$. Then the two equations are solved for a fixed number of iterations, with the same architecture neural net being used to represent \cref{secondsub} in each iteration. However, the network has its own set of weights for each iteration, hence the unrolling of the algorithm. The architecture of the network will be discussed in the next section.

\section{Network Architecture}\label{networks}
\begin{figure*}[!htb]
    \centering
    \includegraphics[width=17.5cm]{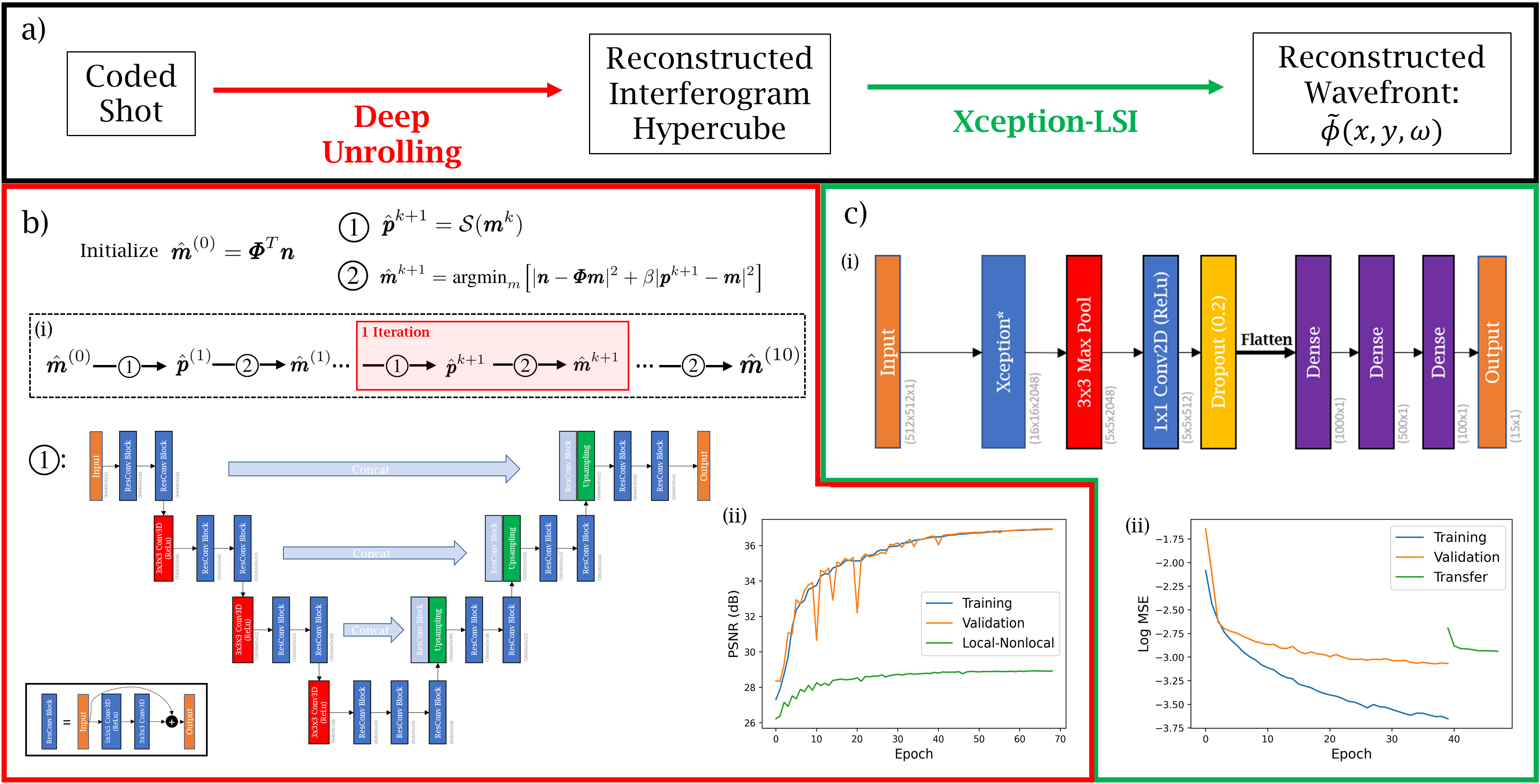}    \caption{A diagram showing the full reconstruction process of the wavefront from the coded shot. a): A flow chart of the reconstruction process. b): (i) The deep unrolling process, where subproblems     {\Large \textcircled{\normalsize 1}} and     {\Large \textcircled{\normalsize 2}} are solved recursively for 10 iterations. Also shown is the neural network structure used to represent $\mathcal{S}(\bm{m}^{k})$. b): (ii) The training curve for the deep unrolling algorithm. Plotted is the training and validation PSNR for the 3D ResUNet prior that was used, as well as the validation score for a local-nonlocal prior. Here is demonstrated the superior power of 3D convolutions in this setting. c): (i) The network design for the Xception-LSI network. The Xception* block represents that the last two layers were stripped from the conventional Xception network. c): (ii) The training curve for Xception-LSI for training and validation sets, with the loss shown in log mean squared error. Also plotted is the validation loss when further training the model on the deep unrolling reconstruction of the data (Transfer).}
    \label{fig:networks}
\end{figure*}

This section contains the architectures of the neural networks that were used. They will be discussed in the order they are used in the reconstruction process, which is displayed in the flow chart of \cref{fig:networks}a. Firstly the deep unrolling algorithm performs reconstruction of the interferogram hypercube from the coded shot, and secondly another network, Xception-LSI, reconstructs the spatial-spectral wavefront from the hypercube.

\subsection{Deep Unrolling Regularizer}
As previously discussed, the neural network, $\mathcal{S}$, represents a regularization term. This means one can exploit prior knowledge about the data to choose a suitable architecture. As will be discussed in the following section, STCs can be described by a correlation between Zernike polynomial coefficients and wavelength. Accordingly, there will likely be strong similarity in spot positions for neighbouring spectral channels. Due to this, an architecture with 3D convolutions was developed, which can exploit these relations. Inspired by recent work in video snapshot compressive imaging\cite{deepunroll3d}, a simplified ResUNet architecture was chosen\cite{resunet}, with the standard 2D convolutions replaced with 3D ones.  We used 10 iterations for our model, as it has been found that adding more than this produces negligible performance gains \cite{spatialspectralunrolling}. A diagram of the network is displayed in \cref{fig:networks}b(i).

\subsection{Xception-LSI}

A wavefront retrieval network was developed that takes a single spectral channel QWLSI interferogram and predicts the spatial wavefront in terms of Zernike coefficients. The network is based on the Xception network\cite{xception}, but as the original 71-layer network is designed for classification, some changes were made to adapt Xception to our application. Firstly the final 2 layers were removed. A max pool layer and a convolutional layer were added to shrink the output in the spatial and spectral dimensions respectively. Dropout was applied before using 3 dense layers with 1000, 500 and 100 nodes using the relu activation function\cite{relu}. The output layer consists of 15 nodes with linear activation, corresponding to the number of Zernike coefficients to predict. We name the network Xception-LSI, and it can be seen in \cref{fig:networks}c(i).

\section{Training data generation}\label{traindata}

To represent the initial pulse, a total of 300 cubes were generated with dimensions $(n_x\times n_y \times n_{\omega}) = (512\times 512 \times 31)$. The data was randomly split at a ratio of $4:1:1$ into training, validation and test sets, respectively. The wavelength range considered was $750-850$ $\SI{}{\nano\meter}$, representing a broadband Ti:Sapphire Laser, giving $\Delta\lambda \approx \SI{3.23}{\nano\meter}$. For each cube, the wavefront for each channel was first initialised to a randomly weighted sum of 15 Zernike basis functions. Then, to simulate an STC, one Zernike function was chosen and was made to vary either linearly or quadratically with frequency. Indeed, common STCs such as pulse front tilt and pulse front curvature can be represented in this way \cite{jolly}. The mean amplitude of this coefficient was also made to be higher. This choice of Zernike coefficients is arbitrary, but allows for a demonstration that the method can identify all Zernike basis functions. The intensity of slices of the cube were set to an image taken of a real laser intensity. 

Each cube was then processed according to \cref{fig:setup}. Firstly it was passed through the QWLSI simulation (see Section IIa), yielding a hypercube of interferograms - these are the training labels for the deep unrolling algorithm. This hypercube was then passed through the SCI simulation, yielding a coded shot - the training data. The wavefront was reconstructed via the process in \cref{fig:networks}a. The interferogram hypercube was reconstructed via deep unrolling, before being passed into the Xception-LSI network to predict the spectral Zernike coefficients. 

The pitch of the lateral shearing interferometer was set to $\Lambda = \SI{80}{\micro\meter}$, and the dispersion of the prism, measured at the camera plane, was set to 1 pixel per channel (each channel having a width of $\SI{3.23}{\nano\meter}$). 

Before being passed through the deep unrolling network, the cubes and coded shots were split spatially into 64$\times$64 oblique parallelepiped patches, allowing for a one-to-one reconstruction between the input and output\cite{spatialspectralunrolling}. The initial learning rate was set to 0.01 and decayed by 10\% every 5 epochs. The total number of epochs was 70, and the batch size was 8. 

The Xception-LSI network was fed individual channels of the ground truth interferogram hypercubes and predicted Zernike coefficient polynomials. Normal random noise ($\mathcal{N}(\mu = 0,\sigma=0.1$)) was applied to the input, to make the model robust to noise produced by the SCI reconstruction. The initial learning rate was set to $10^{-5}$ and decayed by 10\% every 5 epochs. The total number of epochs was 40, and the batch size was 16. Once trained on the ground truth hypercubes, the model was trained on interferogram hypercubes that had been reconstructed by deep unrolling, for a further 8 epochs. The aim of this transfer learning was to allow the network to account for any systematic noise in the SCI reconstruction, resulting in a more accurate wavefront reconstruction. 

\section{Results and Discussion} \label{results}

\begin{figure*}[!htb]
    \centering
    \includegraphics[width=17.5cm]{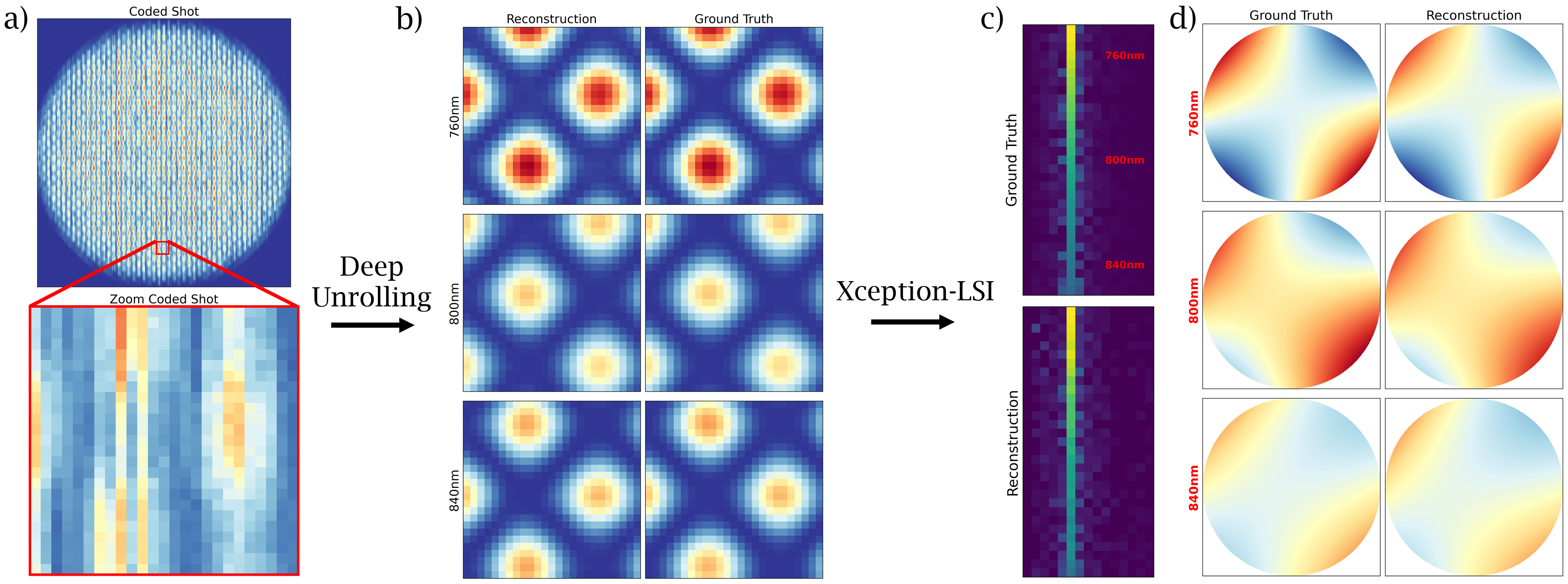}
    \caption{Example results of the reconstruction process. a): An example of the coded shot, along with a zoomed section. b): Deep unrolling reconstruction of the interferogram hypercube in the same zoomed section at different wavelength slices. c): The Xception-LSI reconstruction of the spatio-spectral wavefront displayed in terms of Zernike coefficients, where the x-axis of each plot is the Zernike function, the y-axis is the wavelength and the colour represents the value of the coefficient. d): The spatial wavefront resulting from a Zernike basis expansion of the coefficients in c) at the labelled spectral channels.  }
    \label{fig:finalresult}
\end{figure*}

\subsection{Snapshot Compressive Imaging}
Crucial to this method's success is the SCI reconstruction of the hypercube of interferograms. As can be seen from the green box of \cref{fig:setup}, the image slices are modulated and appear as spot patterns. As a result, the images do not exhibit the same sparsity in e.g. the wavelet domain as most natural images used in SCI research do. Because of this, there was uncertainty in whether it would be possible to recover the cube. 

Here it is demonstrated that it is indeed possible to reconstruct such modulated signals with SCI. The training curve can be seen in \cref{fig:networks}b(ii). Also plotted is the validation loss when a local-nonlocal prior\cite{deep_unrolling}, which is state of the art for natural images, was used. One sees that when both architectures were used with 10 iterations of unrolling, the 3D convolutional model achieved a far superior peak signal to noise ratio (PSNR) of 36 compared to 29. Furthermore, it contains $\sim$45\% less parameters. 

\subsection{QWLSI}

In order to reconstruct the wavefront for a full hypercube, each spectral channel is fed through the network sequentially. After training, the final mean squared error on the ground truth test set was $6.80\times 10^{-4}$. \cref{fig:networks}c(ii) displays the training curve with the training, validation and transfer loss curves. The additional transfer learning proves to be extremely effective in reducing the error of the wavefront predictions when working with reconstructed interferogram hypercubes. The final mean squared error on the reconstructed test set was $9.18\times 10^{-4}$.

\subsection{Hyperspectral Compressive Wavefront Sensing}
An example of the full reconstruction process, from coded shot to spatial-spectral wavefront, is displayed in \cref{fig:finalresult}. It is apparent that the deep unrolling network was able to accurately reconstruct the interferogram hypercube, and the Xception-LSI network was able to reconstruct the wavefront. 

\section{Summary and Outlook}
In this report we have demonstrated the possibility of combining a wavefront sensor with snapshot compressive imaging in order to achieve a single-shot measurement of the spatial-spectral phase. Crucially, it has been shown that SCI has the ability to reconstruct modulated signals, such as those produced by a quadriwave lateral shearing interferometer.  

A natural progression to this study is to realize the results in an experimental setting, where challenges arise from the more complicated dispersion, transfer functions and noise. Other further work could include extending the deep learning LSI analysis to the hyperspectral setting. By passing the network a hypercube of interferograms, rather than individual slices, it may be possible to exploit spectral correlations in order to improve accuracy and detect STCs more easily.   Also, work can be done on testing the model with a more varied set of Zernike polynomials. Finally, there has been recent interest in the possibility of spreading phase contrast imaging to a hyperspectral setting. However, current methods take many seconds to capture a hypercube of phase\cite{PCI}. The proposed method would be able to collect information with higher spectral resolution in a single shot, allowing for dynamic events to be recorded hyperspectrally.

\section*{Acknowledgements}
We would like to acknowledge the useful discussions with Dr Ramy Aboushelbaya and the rest of Professor Peter Norreys' group. 

This work was supported by the Independent Junior Research Group "Characterization and control of high-intensity laser pulses for particle acceleration", DFG Project No.~453619281. We would also like to acknowledge UKRI-STFC grant ST/V001655/1.

\end{document}